\documentclass[12pt]{article}
\usepackage{amsfonts,amssymb} 
\def\be{\begin{eqnarray}}
\def\ee{\end{eqnarray}}
\def\tanc {\rm tanc} 
\def\openone {1\kern -0.36em\llap~1} 
\begin{document}
\phantom{.}

\centerline{\large Umbral Deformations on Discrete Spacetime} 

\phantom{.}

\centerline{ \large Cosmas K Zachos }
High Energy Physics Division,
Argonne National Laboratory, Argonne, IL 60439-4815, USA \\
\phantom{.} \qquad\qquad{\sl zachos@hep.anl.gov}      

           \begin{abstract}
{Given a minimum measurable length underlying spacetime,
the latter may be effectively regarded as discrete, at
scales of order the Planck length. 
A systematic discretization of continuum physics may be effected 
most efficiently through the umbral deformation. 
General functionals yielding such deformations at the level 
of solutions  are furnished and illustrated,
and broad features of discrete oscillations and wave propagation 
are outlined.}
             \end{abstract}

\vskip 0.5cm

\hrule

                   \section{Introduction}
A generic theme of robust arguments \cite{mead} has established an 
expectation  of a fundamental minimum 
measurable length in nature, of the order of  
$l_{_{Planck}} \equiv\sqrt{ \hbar G_N /c^3}\sim 1.6 \cdot 10^{-35}m $, the 
corresponding time for which is $l_{_{Planck}} /c$. 
The essence of such arguments is the 
following (in geometrical Planck units, in which $\hbar$, $c$,  and 
$m_{_{Planck}}\equiv\sqrt{ \hbar c/ G_N }$ are chosen to be 1). 
In a system or 
process characterized by energy $E$, no lengths smaller than $L$ can be
accessed or measured, where $L$ is the {\em larger} of either the 
Schwarzschild horizon radius of the system, $\sim E$, or, for energies smaller 
than the Planck mass, the Compton wavelength of the aggregate process,
$\sim 1/E$.  Since the minimum of max$(E,1/E)$ lies at the Planck 
mass, $E \sim 1$, 
the smallest measurable distance is widely recognized to amount 
to $l_{_{Planck}}$.

Thus, continuum laws in nature are expected to be 
deformed, in  principle, by modifications of $O(l_{_{Planck}})$. Remarkably,
even as something like a fundamental spacetime lattice of spacing 
$a=O(l_{_{Planck}})$  is thus likely to underlie conventional physics, 
continuous symmetries, such as Galilei or Lorentz invariance, can actually 
survive unbroken such  a deformation into discreteness, in a nonlocal,
{\em umbral realization} \cite{LTW}. 

Umbral calculus, pioneered by Rota and associates in combinatorics contexts 
\cite{rota,loeb}, specifies, in principle, how functions of discrete 
variables representing observables get to ``shadow" their continuum 
limit ($a \rightarrow 0$) {\em systematically}, 
and effectively preserve Leibniz's chain rule and the Lie Algebras of the 
difference operators which shadow (deform) the standard differential 
operators of continuum physics.  (For a review useful to physics, see
\cite{levi}.)

Nevertheless, while the continuous symmetries and Lie Algebras of umbrally 
deformed systems might remain identical to their continuum limit, the 
functions of observables themselves are modified, in general, often 
drastically so. Usually, the controlling continuum differential equations of 
physics are discretized \cite{dimakis}. Then, such discrete difference equations 
are solved to yield umbral deformations of the continuum solutions.
However, some complication may be bypassed by umbrally deforming 
the continuum solutions directly. Still, as illustrated below on the 
simplest case of oscillations and wave 
propagation, the modifications may become technically cumbersome in 
extracting the actual physics involved.

In this brief note, an explicit deforming functional 
more tractable than the abstract
umbral deformation definition is articulated, (\ref{neato},\ref{neato2}), 
which is suited to transforming continuum solutions. It is an integral 
functional shadowing the Fourier representation. It should be of utility 
in inferring  wave disturbance propagation in discrete spacetime lattices, 
by evaluating 
the umbral transforms of standard continuum physics quantities. Can generic 
features of discrete spacetime impact wave propagation  at cosmic distances?
We first illustrate the umbral language on harmonic oscillations discretely 
deformed through asymmetric finite differences, $\Delta_{+}$, for simplicity,
so the reader could quickly confirm all statements by standard numerical 
analysis textbook methods (Section 2). We then point out how the umbral 
transform kernel technique proposed may bypass conventional umbral calculus 
complications for more realistic symmetric differences $\Delta_s$, for 
which basic polynomial sets and umbral manipulations are less straightforward 
(Section 3). We conclude by discussing generic features of wave 
propagation on such lattices
at cosmic distances, and the logical possibilities afforded by the 
index of refraction modifications at  $O(l_{-{Planck}})$; we finally propose
conceivable solitonic applications (Section 4).

\section{Overview of the Umbral Correspondence}
For simplicity, consider discrete time, $t=0,a,2a,...,na,...$
Without loss of generality, broadly following the summary review 
of \cite{levi}, consider first the $\Delta_{+}$  
discretization (umbral deformation) of $\partial_t$,
\begin{equation} 
\Delta x(t) \equiv  \frac{x(t+a)-x(t)}{a} ~,
\label{plaindef}
\end{equation}
and whence of the elementary oscillation equation, $\ddot{x}(t)= -x(t)$, namely
\begin{equation} 
\Delta^2 x(t) = \frac{x(t+2a)-2x(t+a) +x(t)}{a^2} = -x(t) ~. \label{diffeqn}
\end{equation}
and investigate the periodicity of its solutions.
Of course, this can be easily solved directly by the textbook Fourier-component
Ansatz $x(t) \propto r^t$,  \cite{benderorszag}, to yield $(1\pm a)^{t/a}$. 
To illustrate the powerful systematics of umbral calculus \cite{levi},
however, we produce and study the solution in that framework, instead.

The umbral framework considers associative chains of operators, generalizing
ordinary continuum functions by ultimately
acting on a translationally-invariant ``Fock vacuum", 1, after 
manipulations to move shift operators to 
the right and have them absorbed in that vacuum.
Using the standard Lagrange-Boole  shift generator 
\begin{equation} 
T\equiv ~ e^{ a \partial_t} ,  \qquad \qquad \hbox{so ~that} \qquad    
T f(t)\cdot 1  = f(t+a) T \cdot 1= f(t+a), 
\end{equation} 
the umbral deformation is   
\begin{equation} 
\partial_t \qquad  \mapsto \qquad  \Delta\equiv \frac{T-1}{a} ~,
\end{equation}
\begin{equation} t \qquad  \mapsto \qquad  tT^{-1} , 
\end{equation}
\begin{equation} t^n \qquad  \mapsto \qquad  (tT^{-1} )^n= 
t(t-a)(t-2a)...(t-(n-1)a) T^{-n} \equiv  [ t ] ^n  T^{-n}  ,\end{equation}
so that $[ t ] ^0 =1$, and, for $n>0$, $[ 0 ] ^n =0$. 
These are called basic polynomials  \cite{rota,levi,dimakis}. 

A linear combination of monomials (a power series representation of a function)
will thus transform umbrally to the same linear combination of
basic polynomials, with the same series coefficients, $f(t) ~~ \mapsto  ~~
f(tT^{-1})$. All observables in the discretized world are thus such 
deformation maps of the continuum observables, and evaluation of 
their direct functional form is in order. 
Below, we will be concluding the correspondence by casually
eliminating translation operators at the very end, through operating on 1, so
that  $f(tT^{-1})\cdot 1$.
 
The umbral deformation relies on the respective umbral entities obeying 
operator combinatorics identical to their continuum limit ($a\rightarrow 0$), 
by virtue of obeying the {\em same Heisenberg commutation relation},
\begin{equation} 
[ \partial_t , t ] ~ = ~ \openone = ~ [ \Delta ,  tT^{-1}  ] ~. 
\end{equation}
(Formally, the umbral deformation reflects (unitary) equivalences of the
unitary irreducible representation of the Heisenberg-Weyl group, 
provided for by the Stone-von Neumann theorem. Here, these equivalences 
reflect the alternate consistent realizations of all continuum 
physics structures through systematic maps such as the one illustrated 
here.)

Thus, e.g., by shift invariance, $T \Delta T^{-1}= \Delta$,  
\begin{equation} 
[ \partial_t , t^n ] =n t^{n-1} \qquad   \mapsto  \qquad   
[ \Delta ,  [t]^n T^{-n}  ] = n [t]^{n-1} ~T^{1-n} , 
\end{equation}
so that, ultimately,  $ \Delta  [t]^n = n [t]^{n-1}$. 

Likewise \cite{rota, ward},
\begin{equation} 
[t]^n T^{-n} [t]^m T^{-m}\equiv 
[t]^n \divideontimes [t]^m T^{-n-m}= [t]^{n+m} T^{-(n+m)},
   \end{equation}
and so forth.
The right member of the equality is the implicit definition of 
the product \cite{ward} defined by dotting on 1,  
$[t]^n \divideontimes [t]^m \equiv [t]^{n+m}$. 

For commutators of associative operators, the umbrally deformed Leibniz 
rule holds \cite{ward,LTW}, 
\begin{equation} 
[\Delta, f(tT^{-1}) g(tT^{-1})] =
[\Delta, f(tT^{-1})]  g(tT^{-1}) + f(tT^{-1}) [\Delta, g(tT^{-1})] ~,
\end{equation}
ultimately to be dotted onto 1.

Now note that, in this case, the basic polynomials $[t]^n$ are just scaled 
falling factorials $a^n (t)_n$,
\begin{equation} 
[t]^n= a^n \frac{(t/a)!}{(t/a-n)!}  ~,
\end{equation}
so that $[-t]^n=(-)^n [t+a( n-1)]^n$. (Furthermore, $[an]^n= a^n n!$ ; 
for  $0\leq m\leq n$, $[t]^m [t-am]^{n-m} =[t]^n$ ; and for    
integers $0\leq m<n$, $[am]^n=0$.  Thus, 
$\Delta^m [t]^n= [an]^m [t]^{n-m} /a^m$.)

The standard umbral exponential is then natural to define as, 
\cite{rota,ward,floreanini},
\begin{equation} 
E(\lambda t,a) \equiv e^{\lambda [t]} \equiv e^{\lambda t T^{-1}} \cdot 1 = 
\sum_{n=0}^{\infty} \frac{\lambda^n}{n!} [t]^n =
\sum_{n=0}^{\infty} (\lambda a)^n   { t/a \choose n } =
(1+\lambda a)^{t/a} ,
\end{equation}
the compound interest formula, with the proper continuum 
limit ($a\rightarrow 0$).
Evidently, since $\Delta \cdot 1 =0$, 
\begin{equation}
\Delta e^{\lambda [t]} = \lambda ~ e^{\lambda [t]} ,
\end{equation}
and, as already indicated, one could have solved this 
equation directly to produce the above $E(\lambda t,a)$ \footnote{ 
There is an infinity of nonumbral extensions of this solution, 
however: multiplying this umbral exponential by an arbitrary periodic function 
$g(t+a)=g(t)$ will not be visible to $\Delta$, and thus will also 
yield an eigenfunction of $\Delta$. Often, as below, such extra 
solutions have a nonumbral, vanishing, continuum limit, or an 
ill-defined one.}.

Serviceably, the umbral exponential $E$ {\em happens to be an 
ordinary exponential}, 
\begin{equation} 
e^{\lambda [t] }= e^{\frac{\ln (1+\lambda a)} {a} t}  .
\end{equation}

The umbral exponential actually serves as the generating function of the umbral 
basic polynomials, 
\begin{equation} 
\frac{\partial^n}{\partial \lambda^n} (1+\lambda   a)^{t/a} 
\Biggr | _{\lambda=0} = [t]^n .
\end{equation}
Conversely, then, this construction may be reversed, by first 
solving directly for the umbral eigenfunction of 
$\Delta$, and effectively defining the umbral 
basic polynomials through the above parametric derivatives, in situations 
where these might be more involved, as in the next section. 

By linearity, the umbral deformation of a power series 
representation of a function formally evaluates to
\be  
f(t) ~~ \mapsto  ~~ F(t)\equiv 
f(tT^{-1})\cdot 1 = f\left( \frac{\partial}{\partial \lambda}\right ) 
 ~(1+\lambda   a)^{t/a} 
\Biggr | _{\lambda=0}, \label{foist} 
\ee 
as a consequence. This may not always be easy to evaluate, but, in fact,
the same argument may be applied to linear combinations of exponentials, 
and hence the Fourier representation, instead,
\be  
F(t) =
\int^{\infty}_{-\infty}\!\! d\tau f(\tau) \int^{\infty}_{-\infty}\frac{dk}{2\pi}
e^{-i\tau k} (1+ika)^{t/a} 
 =  \left (1+ a\frac{\partial}{\partial\tau} \right ) ^{t/a} ~f(\tau) 
 ~\Biggr | _{\tau=0}  . \label{neato}
\ee 
The rightmost equation follows by converting $k$ into 
$\partial_\tau$ derivatives and integrating by parts away from the 
resulting delta function. Naturally, it identifies with the above eqn 
(\ref{foist}) by the (Fourier) identity   
$f(\partial_x)g(x)|_{x=0}= g(\partial_x )f(x) |_{x=0} $. It is up to the 
ingenuity of the particular application of such functionals involved to 
utilize the form whose domain of applicability is best suited for the 
evaluation sought.

It is also straightforward to check that this umbral transform functional 
yields
\be
\partial_t f  ~ \mapsto  ~  \Delta F ; 
\ee 
or to evaluate the umbral transform 
of the Dirac delta function, which amounts to a cardinal sine, or sampling,  
function,  
\be 
\delta(t)~  \mapsto ~  {\sin ({\pi\over 2}  (1+t/a)) \over (\pi (a+t))} ;
\ee 
or to evaluate 
\be
f={1 \over (1-t)}  ~\mapsto ~ F= e^{1/a} a^{t/a}  \Gamma(t/a+1 , 1/a), 
\ee
an incomplete Gamma function, and so on. (For example, discrete integration 
of the above cardinal sine fucntion leads to the umbral transform of the 
step function, a
hypergeometric function $_2F_1$ of imaginary argument.) 
In practical applications, evaluation of umbral transforms of arbitrary  
functions of observables may well be more direct, at the level of 
solutions,  through this deforming functional, eqn (\ref{neato}).

For example, one may evaluate in this way the umbral correspondents of 
trigonometric functions, 
\begin{equation}  
\sin [t] \equiv \frac{e^{i[t]} -e^{-i[t]}}{2i} ~,  \qquad \qquad 
\cos [t] \equiv \frac{e^{i[t]} +e^{-i[t]}}{2} ~,
\end{equation}
so that 
\begin{equation} \Delta \sin [t] =  \cos [t]~, \qquad   \qquad   
\Delta \cos [t]= - \sin [t] .
 \end{equation}

Thus, the umbral deformation of phase-space rotations, 
\begin{equation}
\dot{x}=p, \quad \dot{p}=-x \qquad \mapsto   \qquad 
\Delta x(t)= p(t), \quad  \Delta p(t)=-x(t),   
\end{equation}
readily yields, by directly deforming continuum solutions,  
the oscillatory solutions, 
\begin{equation}
x(t)= x(0) \cos [t] + p(0) \sin [t], \qquad     
p(t)= p(0) \cos [t] - x(0) \sin [t] .
  \end{equation}

As indicated, the umbral exponential being an ordinary exponential; and, by 
\begin{equation}  
(1+ia)= \sqrt{1+a^2} ~ e^{i\arctan (a)} ,
\end{equation}
these solutions are seen to actually amount to discrete phase-space spirals, 
\begin{equation}
\!\! x(t)=(1+a^2)^{t\over 2a} \Bigl ( x(0) \cos (\omega t)  + p(0) \sin (\omega t) 
\Bigr ), \quad 
p(t)=(1+a^2)^{t\over 2a} \Bigl ( p(0) \cos(\omega t)  - x(0) \sin (\omega t) 
\Bigr ),
\end{equation}
with a frequency {\em decreased} from the continuum value 1 to 
\begin{equation} 
\omega = \arctan (a) / a \leq 1  ~, 
\end{equation}
effectively the inverse of the cardinal tangent function. 

That is, for $\theta \equiv  \arctan (a)$, the spacing of the zeros, 
period, etc, are scaled up by a factor of 
\begin{equation} 
\tanc (\theta) \equiv \frac{ \tan  (\theta) } {\theta} \geq 1   ~. 
 \end{equation}
(For complete periodicity on the time lattice, one further needs 
return to the origin in an integral number of $N$ steps, 
thus a solution of $N= 2\pi n  / \arctan a$.)
Example: ~ For $a=1$, the solutions' radius spirals  out as $2^{t/2}$,
while $\omega= \pi/4$, and the period is $\tau=8$.

Note that the umbrally conserved quantity is, 
\begin{equation} 2 {\cal E} = x(t)   \divideontimes  x(t) +p(t) \divideontimes  p(t)=
x(0)^2+p(0)^2=    (1+a^2)^{-t\over a} \Bigl (x(t)^2+p(t)^2\Bigr ),
\end{equation}
($\Delta {\cal E}=0$), with the proper energy as the continuum limit.

\section{More Symmetric Cases}
Unfortunately, the $\Delta_{+}$  of eqn (\ref{plaindef}) is not 
time-reversal odd,
and thus its square is not time-reversal invariant---whence the  awkward
outspiraling of the solutions of the previous section. 
(Its time-reversal conjugate, 
$\Delta_{-}$, would have in-spiraling solutions.) 

Instead, one often 
chooses half the difference of these two delta operators, i.e., 
the time-reversal-odd umbral deformation, 
\begin{equation} 
\partial_t \qquad \mapsto \qquad \Delta_s \equiv \frac{T-T^{-1}}{2a} ~.  
\label{symm}
\end{equation}

The eigenfunctions of ~$\Delta_s E=\lambda E$~ are now two \cite{ltw},
\begin{equation} 
E_{\pm} = 
\Bigl (\lambda a \pm \sqrt{1+ (\lambda a)^2}\Bigr )^{t/a} ;
\end{equation}
one, $E_+$,  going to the exponential in the continuum limit; but 
the other, $E_-$ (``nonumbral"), simply oscillating to zero---an oscillation of 
infinite frequency. 

Now, since $[\Delta_s, 2t/(T+T^{-1}) ]= \openone$,  the basic polynomials, 
$[t]_s^n =\bigl (t~ 2/(T+T^{-1}) \bigr )^n \cdot 1$,
would be harder to evaluate, in principle. Thus, they have been 
evaluated \cite{dimakis} 
from $\Delta_s [t]_s^n =n [t]_s^{n-1} $ instead, 
\begin{equation} 
[t]_s^n=t  \prod_{k=1}^{n-1} \Bigl ( t+a(n-2k) \Bigr ).
\end{equation} 

However, according to the general generating function consideration of the 
previous section, they may alternatively be generated more directly 
from $E_{+}$, 
\begin{equation} 
[t]_s^n =\frac{\partial^n}{\partial \lambda^n} \Bigl (\lambda a +\sqrt{1+ 
(\lambda a)^2}\Bigr )^{t/a} \Biggr | _{\lambda=0} .
\end{equation}
(E.g., the reader may easily check that $[t]_s^3= (t+a)t(t-a)$, etc.)  
Since $E_{-} E_{+}= (-)^{t/a}$, $E_{-}$ generates the reflected 
basis $[t]^n_s (-)^{t/a+n}$. In general, finding the eigenfunctions of
a given difference operator and utilizing them as generating functions
of basic polynomial sets may provide shortcuts in effort.

Rather than using umbral deformations of power series representations for 
functions, however, one may instead infer, mutatis mutandis as in (\ref{neato}),
 umbral transforms  of Fourier representations,
\begin{equation} 
f(t) \qquad \mapsto \qquad F_s(t)= \int^{\infty}_{-\infty}\!\! d\tau f(\tau) 
\int^{\infty}_{-\infty}\frac{dk}{2\pi} e^{-i\tau k} 
\Bigl (ik a +\sqrt{1- 
(k a)^2}\Bigr )^{t/a} ,  \label{neato2}
\end{equation} 
to evaluate umbral deformations for general observables, 
as well as nonumbral ones relying on $E_-$, with a minus sign in the above 
kernel of the deforming functional.

Thus, now there are four solutions to the analog of eqn (\ref{diffeqn}),
\begin{equation} 
(\Delta_s^2 +1) x(t) =0~, 
\label{sdiffeqn}  
\end{equation}
namely
\begin{equation} 
x(t)= \Bigl (\pm ia \pm \sqrt{1-a^2}\Bigr )^{t/a}.  
\end{equation}
Thus the discrete-time solution set  
\begin{equation}
x(t)=(-)^{Nt/a} \Bigl ( x(0) \cos (\omega t)  + p(0) \sin (\omega t)\Bigr ) , 
\quad 
p(t)=(-)^{Nt/a} \Bigl (  p(0) \cos(\omega t)  - x(0) \sin (\omega t) 
\Bigr ), 
\end{equation}
maps onto itself under time-reversal, for integer parameter $N=0,1$.
($\Delta_s x(0)= (-)^N p(0)$.)
This eqn (\ref{sdiffeqn}) 
only connects even points on the time lattice among themselves, and odd ones 
among themselves. Thus, 
all even points on the time lattice behave the same 
for even or odd parameter $N$. However, for the $N=1$ solutions, 
the odd time points hop out of phase by $\pi$ (reflection with respect to 
the origin in phase space), as they are not dynamically linked to the even 
points\footnote{Actually, as above, if $f(t)$ is a solution of 
(\ref{sdiffeqn}),   $g(t)f(t)$ will also be a solution for arbitrary periodic
$g(t+2a)=g(t)$. Thus, even though $(-)^{t/a}$ is one such possible $g(t)$, 
there are even {\em more} solutions with arbitrarily mismatched moduli 
(phase-space radii) and phases between the odd and even sublattices---only 
their frequencies of rotation need be the same. The solution set is 
four dimensional.}, a phenomenon familiar in lattice gauge theory. 

For $N=0$, the frequency is {\em increased} 
over its continuum limit: 
\begin{equation} \omega = \arcsin (a)   / a \geq 1. 
 \end{equation}
For $N=1$, the arcsine effectively advances by $\pi$ 
and the frequency has an additional component of $\pi /a$. Thus, these 
nonumbral solutions collapse to 0 in the continuum limit.

The conserved energy is more conventional, 
\begin{equation}   
2{\cal E}_s  = x(t)^2+p(t)^2.
\end{equation}

This reversal-odd difference operator (\ref{symm}) is the 
one to be considered in wave propagation in the next section, to avoid 
presumably unphysical exponential amplitude 
modulations, growths or dwindlings, peculiar to the asymmetric derivative seen
in the previous section.

\section{Wave Propagation Outlines} 
Given the features of discrete oscillations outlined, simple
plane waves in a positive or negative direction $x$ would 
obey an equation of the type \cite{floreanini,ltw}, 
 \begin{equation}   
( \Delta_x^2 -\Delta_t^2)~  F_s=0, 
\end{equation}
with the symmetric difference operators of the type (\ref{symm}) on a time 
lattice with 
spacing $a$, and an $x$-lattice of spacing $b$, respectively, not necessarily 
such that $b=a$ in some spacetime regions. 

For generic frequency, wavenumber and velocity,
the basic right-moving waves $ e^{i(\omega t - kx)}$ have phase velocity 
\be    
 v(\omega, k) = \frac{\omega}{k} ~ \frac{a \arcsin (b) }{b \arcsin (a) }~~,  
\ee 
that is to say, the effective index of refraction in the discrete medium is
$(b \arcsin (a))/(a \arcsin (b)) $, so modified from 1 by $O(l_p)$. 

Small inhomogeneities of $a$ and $b$ in the fabric of spacetime over large 
regions could yield interesting frequency shifts in the index of refraction,
and thus, e.g., whistler waves over cosmic distances. It might be worth 
investigating application of the umbral deformation functional (\ref{neato2})
on such waves, to access long range effects of microscopic qualifications 
of the type considered.

A further, more technically challenging application of the umbral tranforms 
proposed might
attain significance on nonlinear, solitonic phenomena, such as, e.g., 
the one-soliton solution of the continuum Sine-Gordon equation,
\be
(\partial_x^2 - \partial_t^2) f(x,t)= \sin (f).
\ee
The corresponding umbral deformation of the equation itself
would now also involve a deformed potential 
$\sin (f(x~ {2\over T_x+T_x^{-1}}, t~ {2\over T_t+T_t^{-1}})) \cdot 1$ on the 
right-hand side, for the $\Delta_{s}$ deformation---and  \linebreak  
$\sin (f(xT^{-1}_x, t T^{-1}_t)) \cdot 1$ for the $\Delta_{+}$ deformation.
 
As illustrated here, rather than solving difficult nonlinear difference 
equations, one may instead infer the umbral transform that, e.g.,  
the continuum one-soliton solution maps to,
\begin{equation} 
F_s=\int^{\infty}_{-\infty}\!\!  \frac{d\chi d\tau dp dk   }{\pi^2} 
\arctan \left ( m e^{\frac{\chi-v\tau}{\sqrt{1-v^2}}}\right ) ~
e^{-i\chi p -i\tau k} 
\Bigl (ip b +\sqrt{1- (pb)^2}\Bigr )^{x/b} 
\Bigl (ik a +\sqrt{1- (k a)^2}\Bigr )^{t/a}.  
\end{equation} 
For the $\Delta_{+}$ deformation,
one would have, instead,
\begin{equation} 
F_{+} =\int^{\infty}_{-\infty}\!\!  \frac{d\chi d\tau dp dk   }{\pi^2} 
\arctan \left ( m e^{\frac{\chi-v\tau}{\sqrt{1-v^2}}}\right ) ~
e^{-i\chi p -i\tau k} 
\Bigl (ip b + 1\Bigr )^{x/b} 
\Bigl (ik a +1\Bigr )^{t/a}.  
\end{equation} 
Closed form evaluation of this integral appears complicated, but 
it could be plotted numerically and the 
representation could yield qualitative asymptotic insights on the 
$O(l_{Planck})$ modifications of 
such umbral solitons.

Likewise, the analog integrals with the continuum 
KdV soliton $f(x,t)= \frac{v}{2} \hbox{sech} ^2 (\frac{\sqrt{v}}{2} (x -vt))$ 
as input,
\begin{equation} 
F_s=\int^{\infty}_{-\infty}\!\!  \frac{d\chi d\tau dp dk  ~v }{8\pi^2} 
\hbox{sech} ^2 \left (\frac{\sqrt{v}}{2} (\chi-v\tau)\right ) ~
e^{-i\chi p -i\tau k} 
\Bigl (ip b +\sqrt{1- (pb)^2}\Bigr )^{x/b} 
\Bigl (ik a +\sqrt{1- (k a)^2}\Bigr )^{t/a},  
\end{equation} 
and, for the $\Delta_{+}$ deformation,
\begin{equation} 
F_{+} =\int^{\infty}_{-\infty}\!\!  \frac{d\chi d\tau dp dk ~v  }{8\pi^2} 
\hbox{sech} ^2 \left (\frac{\sqrt{v}}{2} (\chi-v\tau) \right ) ~
e^{-i\chi p -i\tau k} 
\Bigl (ip b + 1\Bigr )^{x/b} 
\Bigl (ik a +1\Bigr )^{t/a},  
\end{equation} 
 could be plotted numerically and 
compared to the Lax pair integrability machinery of Ref \cite{levi}, or the 
results on a variety of discrete KdVs in ref \cite{schiff}.
These questions are left for a forthcoming study. The ingenuity of further
particular applications of the deforming functionals proposed here is 
left to the reader.

\noindent{\sl \small 
This work was supported by the US Department of Energy, 
Division of High Energy Physics, Contract DE-AC02-06CH11357.
Helpful discussions with D K Sinclair, J Uretsky, and D Fairlie 
are gratefully acknowledged.}


\begin{thebibliography} {99} 
\bibitem{mead}     C Mead,  Phys Rev {\bf 135} (1964)  B849 - B862 ;\\ 
L Garay, Int J Mod Phys  {\bf A10} (1995) 145-166 [gr-qc/9403008];\\ 
X Calmet, M Graesser, and S Hsu,  Phys Rev Lett {\bf 93} (2004)  211101; \\ 
S Liberati,  S Sonego, and M Visser, Phys Rev {\bf D71} (2005)  045001. 
\bibitem{LTW}  D Levi, P Tempesta, and P  
Winternitz,  Phys Rev {\bf D69} (2004) 105011 [hep-th/0310013];\\
D Levi, J Negro, and M del Olmo,  J Phys {\bf 34} (2001)  2023--2030.  
\bibitem{rota} G-C Rota, {\em Finite Operator Calculus} (Academic Press, 1975). 
\bibitem{loeb} A Di Bucchianico and D Loeb, Electron J Combin {\bf DS3} (1995) 
[update 4/2000].
\bibitem{levi} D Levi and P Winternitz, J Phys {\bf A39} (2006) R1-R63. 
\bibitem{dimakis} A Dimakis, F M\"{u}ller-Hoissen, T Striker, Jou 
Phys {\bf A 29} (1996)  6861-6876.  
\bibitem{benderorszag} C Bender and S Orszag, {\em Advanced Mathematical 
Methods for Scientists and Engineers} (Mc Graw-Hill, 1978). 
\bibitem{ward} R Ward, Phys Lett {\bf A165} (1992) 325-329; \\
Y Bouguenaya and D Fairlie, J Phys {\bf A19} (1986) 1049-1053.
\bibitem{floreanini} R Floreanini and L Vinet, J Math Phys {\bf 36} 
(1995) 7024-7042; \\ D Levi, J Negro, and M del Olmo,  
J Phys {\bf 37} (2004)  3459-3473. 
\bibitem{ltw} D Levi, P Tempesta, and P Winternitz, J Math Phys {\bf 45} 
(2004) 4077-4105. 
\bibitem{schiff} J Schiff, Nonlinearity {\bf 16} (2003) 257-275  [nlin/0209040].
\end{thebibliography}
\end{document}